\documentclass[journal,a4paper,twoside]{IEEEtran}

\usepackage[english]{babel}
\usepackage[T1]{fontenc}
\usepackage{comment}
\usepackage{silence}

\usepackage[hyphens,lowtilde]{url}
\usepackage{eucal}
\usepackage{verbatim}
\usepackage{epsfig}
\DeclareGraphicsExtensions{.eps}

\usepackage[utf8]{inputenc}
\usepackage{tikz}
\usepackage{filecontents}
\usepackage{cite}
\graphicspath{{.}{./fig/}}

\usepackage{amsfonts}
\usepackage{amsmath}
\usepackage[makeroom]{cancel}
\usepackage{algorithm}
\usepackage{color}
\usepackage{array}
\usepackage{bm}
\usepackage{enumitem} 
\usepackage{booktabs} 
\usepackage[flushleft]{threeparttable} 
\usepackage{multirow}
\setlistdepth{9}

\ifCLASSOPTIONcompsoc
  \usepackage[caption=false,font=normalsize,labelfont=sf,textfont=sf]{subfig}
\else
  \usepackage[caption=false,font=footnotesize]{subfig}
\fi

\usepackage{placeins}
\usepackage{dblfloatfix}

\newlist{deepEnumerate}{enumerate}{9}
\setlist[deepEnumerate,1]{label=\arabic*.}
\setlist[deepEnumerate,2]{label=\alph*)}
\setlist[deepEnumerate,3]{label=\roman*.}
\setlist[deepEnumerate,4]{label=\Alph*.}
\setlist[deepEnumerate,5]{label=\Roman*.}
\setlist[deepEnumerate,6]{label=\arabic*.}
\setlist[deepEnumerate,7]{label=\alph*)}
\setlist[deepEnumerate,8]{label=\roman*.}
\setlist[deepEnumerate,9]{label=\Alph*.}

\renewcommand\IEEEkeywordsname{Index Terms} 

\begin{document}

\title{Design Approach for Additive Manufacturing in Spare Part Supply Chains}

\author{Filipe M. de Brito,
        Gélson da Cruz Júnior,
        Enzo M. Frazzon,
        João P. Basto,
        Symone G.\ S.\ Alcalá
\thanks{Filipe M. de Brito and Gélson da Cruz Júnior are with the School of Electrical, Mechanical and Computer Engineering, Federal University of Goiás, Goiânia, 74605-010, Goiás, Brazil, and also with INESC P\&D Brasil, São Paulo 11055-300, Brazil (e-mails: filipe.marinho.brito@gmail.com; gcruzjr@ufg.br).}
\thanks{Enzo M. Frazzon is with the Graduate Program in Production Engineering, Department of Production Engineering and Systems, Federal University of Santa Catarina, Florianópolis, 88040-900, Brazil, and also with INESC P\&D Brasil, São Paulo 11055-300, Brazil (e-mail: enzo.frazzon@ufsc.br).}
\thanks{João P. Basto is with the Department of Industrial Engineering and Management, Faculty of Engineering, University of Porto, Porto, 4200-465, Portugal, and also with the INESC TEC - Technology and Faculty of Engineering of the University of Porto, Porto, 4200-465, Portugal (e-mail:joao.p.basto@inesctec.pt).}
\thanks{Symone G.\ S.\ Alcalá is with the Production Engineering, Faculty of Sciences and Technology, Federal University of Goiás, Aparecida de Goiânia, 74968-755, Goiás, Brazil, and also with INESC P\&D Brasil, São Paulo 11055-300, Brazil (e-mail: symone@ufg.br).}}

\IEEEpubidadjcol


\maketitle

\begin{abstract}
In the current industrial revolution, additive manufacturing (AM) embodies a promising technology that can enhance the effectiveness, adaptability, and competitiveness of supply chains (SCs). Moreover, it facilitates the development of distributed SCs, thereby enhancing product availability, inventory levels, and lead time. However, the wide adoption of AM in industrial SCs creates various challenges, leading to new difficulties for SC design. In this context, this paper proposes a new design approach to AM SCs using optimization methods. More specifically, the proposed approach, comprising the $p$-median and mixed-integer linear programming models, considers the decision of deploying productive resources (3D printers) in specific locations of generic spare part SCs. The approach was evaluated in a real-world use case of an elevator maintenance service provider. The obtained results demonstrated the promising capabilities of the proposed design approach in managing the challenges arising from the forthcoming widespread use of 3D printers in manufacturing SCs.
\end{abstract}


\IEEEpubid{\begin{minipage}{\textwidth}\ \\ \\ \\ \\ \\ \\ [12pt]
  \fbox{\begin{minipage}{\textwidth}\copyright 2020 IEEE. Personal use of this material is permitted. Permission from IEEE must be obtained for all other uses, in any current or future media, including reprinting/republishing this material for advertising or promotional purposes, creating new collective works, for resale or redistribution to servers or lists, or reuse of any copyrighted component of this work in other works. DOI: 10.1109/TII.2020.3029541\end{minipage}}
\end{minipage}} 

\IEEEpeerreviewmaketitle

\section{Introduction}
\label{IntPaper}

The fourth industrial revolution, also called Industry 4.0, denotes a new industrial era that employs emerging digital technologies to increase quality, productivity, flexibility, and efficiency in production systems\cite{frank2019industry}.

New technologies, emerging from this revolution, are changing the design and operation of supply chains (SCs). SC design should carefully consider all entities involved entities (e.g., suppliers, distribution centers, and transportation) to achieve optimal performance \cite{garcia2015supply}. Moreover, different factors such as transport activities, scale, and complexity should be considered. The location of the facilities is particularly relevant when the company operates on a global scale \cite{garcia2015supply}.

In this context, the potential advantages of additive manufacturing (AM) are a high customization level, flexible product design, and reduced complexity of SCs \cite{bogers2016additive}). These characteristics enable manufacturing systems, SCs, and products to be improved, and more complex parts can be produced at a lower lead time and in a more cost-effective manner \cite{trancoso2018simulation}. In generic spare parts SCs, the use of AM has significant potential that should be explored \cite{khajavi2014additive}. This has been reinforced by studies \cite{frazzon2018hybrid,frazzon2018data}, which affirms that current industrial facilities will evolve into smart factories and encounter the challenges of shorter product life cycles, mass customization, and an increasingly intense global competition. AM enables the production of old spare parts, eases the management of stock levels of products with a high level of customization, and decentralizes production, allocating it closer to the final customer. A major shift is the move from centralized to decentralized SCs \cite{bogers2016additive}, with a focus on localization and accessibility to increase the availability of parts in challenging locations of the SCs. 

The integration of AM into the SC design may be considered a strategic decision that is a part the 4$^{th}$ Industrial Revolution. The advantages provided by these technologies have demonstrated the potential to significantly affect SCs, manufacturing systems, and products, enabling more complex parts to be manufactured with reduced cost and lead time \cite{trancoso2018simulation}.

Studies on the opportunities and effects of AM in SCs \cite{holmstrom2010rapid,khajavi2014additive,emelogu2016additive,khajavi2018additive} and the implementation of AM and make-or-buy decisions \cite{mellor2014additive,ruffo2006cost,khajavi2018additive} indicated that the use of AM technology may have an effect on different members of an SC (e.g., suppliers and customers) and the potential to enable a distributed production of (spare) parts in a make-to-order (MTO) strategy. In \cite{holmstrom2010rapid}, Holmström et al. indicated that AM can simplify SCs by narrowing and shortening them. This is because they provide the opportunity to integrate additional functionality into products and to optimize products for their function \cite{holmstrom2010rapid}. Hence, this reduces the number of sub-components and suppliers required. In \cite{berman20123}, Berman suggested that transferring small-batch AM production back from low- to high-wage countries may reduce the necessity for manual labor. Product lead time, demand, and the ratio between the unit production costs of AM and traditional manufacturing are key cost parameters to analyze the make-or-buy decisions and determine the economic feasibility of AM \cite{emelogu2016additive}.

\IEEEpubidadjcol

This is most relevant for firms that offer customized products down to a lot size of one. In addition, according to \cite{khajavi2018additive}, depending on the SC configuration, a positive relationship between multi-part production capability and feasibility exists on AM-enabled spare part SCs with hub configurations. Moreover, operations of companies can become more agile with AM \cite{vinodh2009agility}, primarily because of the ability of the technology to alter product designs rapidly. Customers of AM products can benefit from higher service levels, as production may be decentralized and, thus, occur closer to the customer \cite{holmstrom2010rapid,khajavi2014additive}.

In the outlined scenario, optimization techniques can contribute to the new SC design with the continuous adoption of AM. This can be achieved by developing models (such as mathematical models \cite{hashimoto2016supply},\cite{emelogu2016additive}) to reduce the total operational costs, even for large SCs. Optimization processes can aid in the decision to deploy productive resources (3D printers) in specific locations of a generic spare part SC that satisfies a predefined lead time (i.e., the total time of production and delivery of a part to the client).

This paper proposes a design approach based on a mixed-integer linear programming (MILP) optimization model, which supports the decision to deploy 3D printers in specific locations of a generic spare part SC that operates under an MTO strategy, resulting in a completely new SC design. Specifically, a real-world use case of an elevator maintenance service provider is addressed herein. The main objective is to decide the number of 3D printers allocated to each internal facility (IF) (to produce spare parts) and its location in addition to the supplier of each spare part. The proposed approach minimizes the total SC costs and can be applied to larger SCs through the use of the $p$-median model \cite{reese2006solution} to cluster demand input and reduce the search space for locating production centers. Moreover, it uses the location-allocation model to explore trade-offs between a make-or-buy logic applied to the 3D printing or external acquisition of a part. Three scenario tests with different types of demand are addressed to demonstrate the promising capabilities of the proposed methods in addressing new design challenges emerging from the widespread use of 3D printers in manufacturing SCs.

This paper is organized as follows. Section \ref{sec:key_concepts} presents background information on the $p$-median, location-allocation, and MILP approaches. Section \ref{sec:case_study} introduces the real-world use case. Section \ref{sec:proposed_methodology} details the proposed approach to design an optimal SC using mathematical models. Section \ref{sec:experimental_results} presents and evaluates the experimental results of the proposed approach thorough the use case. Finally, Section \ref{sec:conclusions} presents the concluding remarks and future studies.

\section{Background}
\label{sec:key_concepts}

In this section, the main concepts of the $p$-median, location-allocation, and MILP approaches are described.

\subsection{P-median}
\label{subsec:pmedian}
The $p$-median algorithm solves the problem of locating $p$ facilities to minimize the sum of distances from each node (client) to the nearest facility. This problem was initially addressed by de Fermat, who attempted to determine a median point in a triangle (three points in a plane) to minimize the sum of the distances from each point to the median point. Later, Weber included weights to each of the three points to reproduce customer demands \cite{weber1929theory}. Subsequently, this problem was improved to a multi-facility approach to determine the median of $n \ge 3$ points, which generalizes to the question of selecting $p > 1$ medians at continuous locations in a Euclidean plane \cite{reese2006solution}.

Hakimi addressed similar problems to calculate medians on a graph \cite{hakimi1964optimum}. He generalized the absolute median to determine $p$-medians on a graph to minimize the sum of the weighted distances. The main contribution of this solution was that the points could be placed anywhere along the edges of the graph. Solutions with $p$ vertices are called $p$-medians of the graph \cite{reese2006solution}. Researchers have used the $p$-median model in several problems, such as in the spatial distribution of maternity hospitals \cite{baray2013optimizing}, in the location of public schools \cite{menezes2014locating}, and classic facilities-location problems \cite{adasme2018p}. The main advantages of the $p$-median algorithm include the following: the flexibility to manage a large variety of input formats (e.g., square-symmetric and square-asymmetric) \cite{kohn2010p}, and many facility location problems can be adapted as $p$-median problems through data processing \cite{cadenas2011}.

\subsection{Location-allocation}
\label{subsec:location_allocation}

In the location-allocation problem, several facilities are placed between many customers to minimize the transportation costs (or the total SC cost) from the facilities to the customers \cite{wen2014alpha}. This problem was originally proposed by Cooper \cite{cooper63} and has been studied by many researchers. Real-world facility location-allocation problems lack information on customers' demands \cite{wen2014alpha} and are frequently uncertain or frequently change (that is, demands, costs, locations, and other factors change over time). In general, location-allocation problems employ solvers to determine optimal locations for facilities that will support the demand for a given set of customers \cite{cplex2009v12}. These solvers can assign the demand to facilities, considering factors such as the number of facilities available, their costs, and the maximum capacity from a facility. For example, in \cite{mogale2018}, a location-allocation problem using mathematical models was addressed to minimize the cost of a food grain SC and the lead time, as well as decisions such as location/allocation, inventory, capacity, and transportation.

\subsection{Mixed-Integer Linear Programming}
\label{subsec:milp}
MILP is a mathematical model with linear constraints that requires a subset of variables with integer values \cite{karlof2005integer}. MILP methods have been widely used by many researchers in business and engineering. This is primarily because MILP is a solver based on linear programming, and its modeling is flexible \cite{vielma2015mixed}. MILP problems are based on the branch-and-bound method through linear-programming relaxation \cite{efroymson1966branch}. They have been applied in various non-deterministic polynomial time (NP)-complete problems (that is, a subset of decision problems whose solutions can be verified in NP when all other problems in the NP set can be transformed into a problem $p$ in polynomial time), such as the SC design problem. For example, an MILP model for a liquefied natural gas SC planning problem, considering a given time horizon, was proposed in \cite{Sangaiah2019} to minimize the costs of the vendor. Moreover, an MILP problem was addressed in \cite{Ding14} to determine a demand response energy management scheme for industrial facilities, where the main objective was to minimize the energy cost of the facilities.

\setcounter{figure}{1}
\begin{figure*}[!b]
\centering
\par\vspace{-0.4cm}\par
\includegraphics[width=0.7\textwidth]{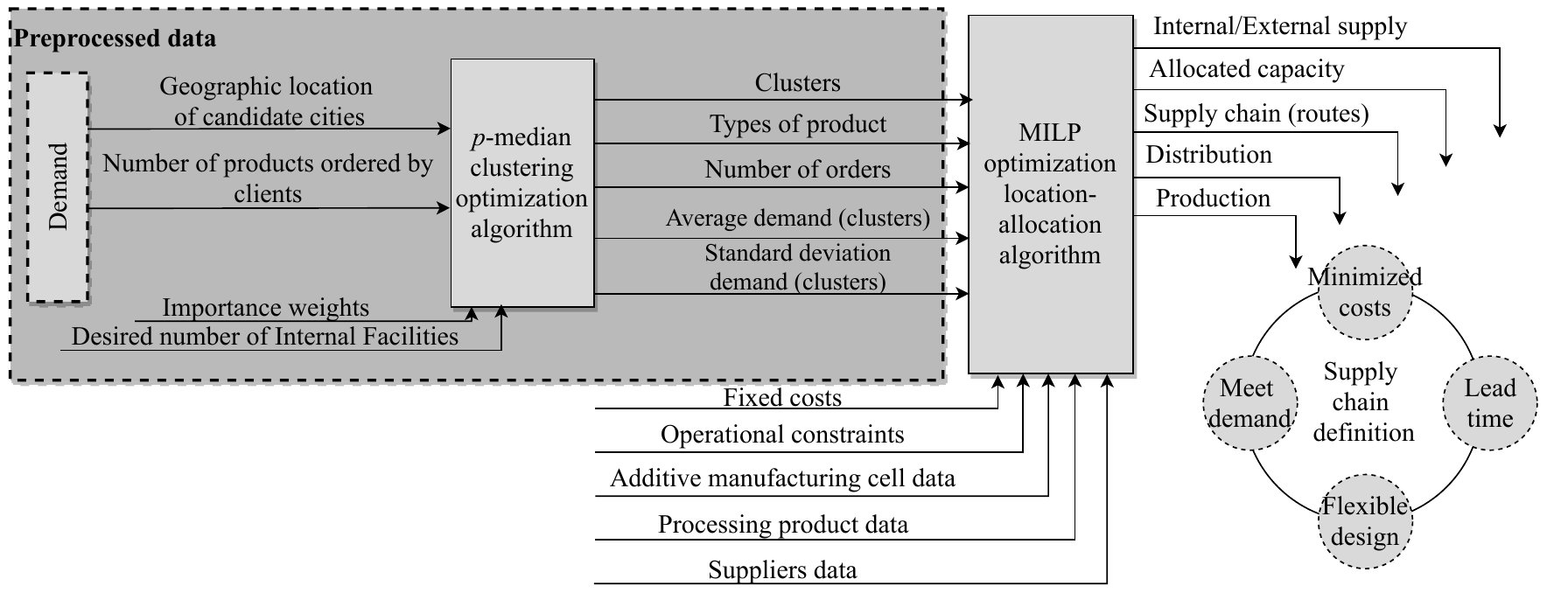}
\par\vspace{-0.4cm}\par
\caption{Optimization procedure flowchart.}
\label{fig:flowchart}
\end{figure*}

\setcounter{figure}{0}
\begin{figure}[!htb]
\centering
\par\vspace{-0.3cm}\par
\includegraphics[width=0.5\textwidth]{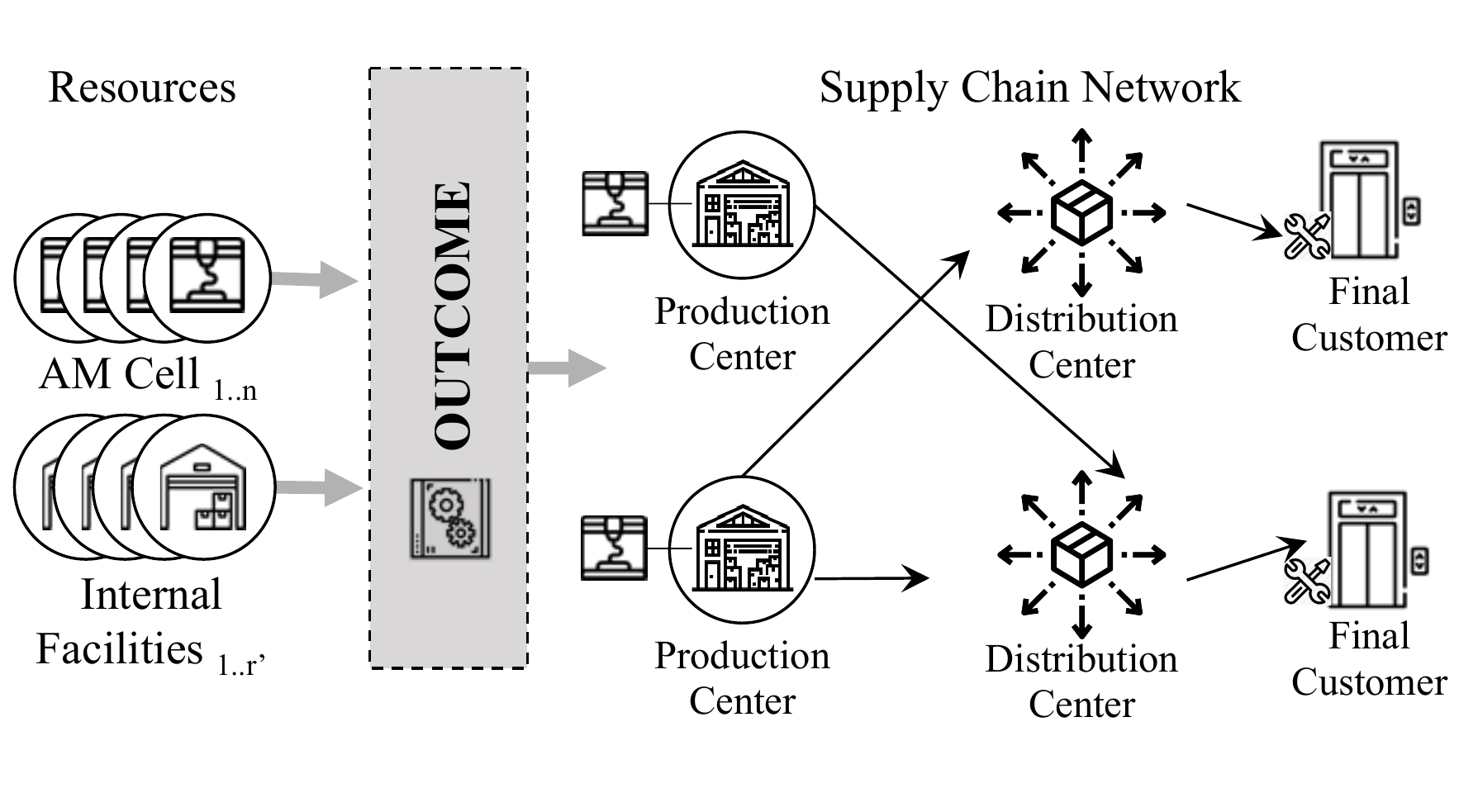}
\par\vspace{-0.7cm}\par
\caption{Overview of the spare part SC design problem.}
\label{fig:describe_the_problem}
\par\vspace{-0.5cm}\par
\end{figure}

\section{Problem Description}
\label{sec:case_study}
In this paper, a real-world industrial case study of a provider of elevator maintenance services in Brazil is addressed. Its branch concentrates on elevator manufacturing and service. The company supplies, within Brazil to other South American countries, components for maintenance services on their elevators and escalators. It aims to employ the AM process in the production of spare parts (using 3D printers) in some Brazilian cities to decrease its costs.

Currently, the company operates with four models of 3D printers: Makerbot Replicator 2, Makerbot Z-18, Cliever Cl2-Pro, and Mankati Fullscale XT Plus. They employ a fused deposition modeling process. The printed spare parts are small polymer parts, and the used printing materials are polylactic acid (PLA), flexible PLA, nylon, acrylonitrile-butadiene-styrene, and polyethylene terephthalate glycol.

In the company, the low lead time and the cost to produce spare parts must be considered to improve its manufacturing and logistic processes. For this, in a city, an IF selected to be an AM cell (also known as a production center (PC)) will supply spare parts to other IFs, also known as distribution centers (DCs). Thus, the problem involves the decision of distributing AM cells (with 3D printers) in specific locations or cities to design a spare part SC (Fig. \ref{fig:describe_the_problem}).

Therefore, the main objective of this paper is to develop a model that is strategic and optimizes the total costs of producing and distributing spare parts at a lower cost for a given lead time. The main decisions involved in this problem are as follows: which cities should be selected to contain an IF given the demand to be satisfied? Which cities (or IFs) will operate as DC or PC? Who should be the supplier for each part: a PC or an external supplier (ES)? and How many 3D printers should be placed with each PC? The optimizer must minimize the total cost of the SC, which involves the annual running costs of the SC, acquisition of spare parts, production, transportation, and delivery costs. 

\section{Formulation and Optimization Approach}
\label{sec:proposed_methodology}

To address the decision of deploying productive resources involved in the SC design problem, the proposed approach splits the problem into two phases: a \textit{pre-processing phase}, by using the classical $p$-median modeling, and an \textit{optimization phase}, by using location-allocation modeling (Fig. \ref{fig:flowchart}).

Firstly, the $p$-median optimization model is employed to cluster demands. That is, for a given number of IFs, the $p$-median obtains clusters of customers based on their delivery time and transportation costs. Then, for each cluster, an optimal IF is chosen based on the demand placed by each customer and its location. In this phase, some assumptions are considered: demand points are considered at regular intervals along with a 2D geographic object (the map), and a demand point is a possible location for an IF. In this phase, the outputs are a set of candidate IFs that best supply each geographic region.

Subsequently, in the second phase of the proposed approach, the location-allocation optimization model is applied to answer the following: which IFs should be PCs, what number of 3D printers in each PC should supply the demand and the supply relations between PC and IFs, and whether the ES or IFs supply the parts. The following subsections present the objective function, constraints, inputs, setups, model, parameters, decision variables, and outputs for each phase.

\subsection{Pre-processing phase}

As described above, this phase performs a demand clustering using a $p$-median algorithm \cite{cplex2009v12} to determine the candidate IFs that supply each geographic region best.

\subsubsection{Objective function}
\label{subsec:objective_function}
The objective function of the $p$-median model minimizes the total cost of SC as follows:
\begin{align*}
{\text{min}} 
\sum_{r \in R, f \in F} Demand_{f} * TotalCost_{rf} * Route_{rf},
 \stepcounter{equation}\tag{\theequation}\label{eq:1}
\end{align*}
where $Demand_{f}$ is the number of products ordered by client $f$, $r$ is an IF, $TotalCost_{rf}$ is the normalized total cost of $r$ supplying $f$, and $Route_{rf}$ is a binary variable (1 if an IF $r$ supplies a client $f$, and 0 otherwise).

\subsubsection{Constraints}
The proposed model considers the following problem constraints: $i)$ a client is supplied by one and only one IF (meaning that the order is not split, and the problem is not a transportation problem) as:
\begin{align*}
\sum_{r \in R} Route_{rf}=1,\forall f \in F;
\stepcounter{equation}\tag{\theequation}\label{eq:2}
\end{align*}
$ii)$ the number of IFs must be equal to $p$ as follows:
\begin{align*}
\sum_{r \in R} OpenIF_{r}=p,
\stepcounter{equation}\tag{\theequation}\label{eq:3}
\end{align*}
where $OpenIF_{r}$ is a binary variable (1 if an IF $r$ should be opened, and 0 otherwise), and $p$ is the number of IFs; and 
$iii)$ only one route from an IF $r$ to a client $f$ can exist, irrespective of whether an IF exists:
\begin{align*}
 Route_{rf}\leq OpenIF_{r}, \forall r \in R,\forall f \in F.
 \stepcounter{equation}\tag{\theequation}\label{eq:4}
\end{align*}

\subsubsection{Inputs}
The selected inputs are the desired number of IFs, geographic locations of candidate cities, number of products ordered by client, and weight (at the interval of $[0,1]$) that balances the contribution of travel time and transportation distance, where larger weight values of one produce a lower contribution of the other.

\subsubsection{Setup}
\label{subsec:setup}
This step focuses on the cost normalization, for which a unity-based normalization is used to perform a nondimensionalization of lead time and distance to enable weighting importance between these parameters. Moreover, a web service through Google's Matrix API\footnote{\url{https://developers.google.com/maps/documentation/distance-matrix/client-library}} is used to obtain the geographic locations of IFs, road trip distance, and travel time between locations. The total cost to supply from $i$ location to $j$ location is calculated as
\begin{align*}
TocalCost_{ij} = distanceWeight * \left(\frac{distance_{ij}}{maxDistance}\right) 
\\ + timeWeight * \left(\frac{travelTime_{ij}}{maxTime}\right),
\stepcounter{equation}\tag{\theequation}\label{eq:5}
\end{align*}
where $timeWeight$ and $distanceWeight$ are complementary unity-based normalized weights, $distance_{ij}$ is the distance value (in meters) of locations $i$ and $j$ for road trips, $maxDistance$ is the largest distance value in the set of all distance values, $distanceWeight$ is a weighting importance to travel distances of $i$ and $j$, $timeWeight$ is a weighting importance to travel times of $i$ and $j$, $travelTime_{ij}$ is the travel time value (in seconds) from location $i$ to $j$ in a road trip, and $maxTime$ is the largest time value (in seconds).

\subsubsection{P-median model parameters}
$r$, $f$, $p$, $TocalCost_{rf}$, and $Demand_{f}$.

\subsubsection{Decision variables}
$OpenIF_{r}$ and $Route_{rf}$.

\subsubsection{Outputs}
\label{subsec:preprocessing output}
The outputs are a set of candidate IFs and their respective clients. The $p$-median algorithm groups the orders in each IF 
to perform a post-processing calculation to determine the number of orders, average demand, and standard deviation demand of a candidate IF. The inputs are sorted by IFs and their clients, and the main aim of the post-processing step is to apply a filter, while prioritizing optimal locations to allocate IFs, in addition to processing all demands. This procedure focuses on reducing transport costs (in this paper, they are represented by the normalized weighted values of $distance_{ij}$ and $travelTime_{ij}$) whose component is often the largest in calculating the total cost of logistics problems. Hence, the $p$-median algorithm considers only data of demand and transport costs; therefore, it cannot solve the SC design.

In this step, the outputs are the geographic locations of candidate IFs and clients, types of products, groups of clients supplied by an IF, demands sorted by IFs and clients, number of orders of each IF, average demand of candidate IFs, and the standard deviation demand of candidate IFs.

\subsection{Optimization phase}
In this phase, the location-allocation algorithm is applied \cite{cplex2009v12}. It uses an MILP approach to determine which candidate IFs must become PCs and the number of 3D printers that should be allocated to each PC to ensure the capacity to supply the demands. Meanwhile, the pre-processing phase evaluates the demand and geographical location, and the optimization phase considers annual fixed costs, production capacity, and ES information.

\subsubsection{Objective function}
\label{subsec:opt_objective_function}
The objective function of the optimization model minimizes the total cost of the SC as follows:
\begin{align*}
{\text{min}} \sum_{r \in R} x_{r} * (FC^R + n_r * FC^P) +
\\\sum_{r,r' \in R, p\in P} y_{r'rp} * \left( uc_{rp} * D_{r'p} + dc_{rr'}*no_{r'p} + ioc_{rp}*no_{r'p}\right) 
\\+ \sum_{s \in S, r' \in R, p \in P} z_{sr'p}*\left(price_{sp}*D_{r'p} + sdc_{sr'}*no_{r'p}\right),
\stepcounter{equation}\tag{\theequation}\label{eq:6}
\end{align*}
where $x_{r}$ is a binary variable (1 if an IF $r$ exists, and 0 otherwise); $FC^{R}$ is the annual fixed cost of an IF $r$; $n_r$ is the number of allocated 3D printers at each PC $r$; $FC^{P}$ is the annual fixed cost of holding a 3D printer; $y_{rr'p}$ is a binary variable (1 if a route exists from IF $r$ to IF $r'$ to deliver a part $p$); $uc_{rp}$ is the unit cost of production of part $p$ in IF $r$, which involves an average hourly cost of human resources (\$/h), employed labor, feedstock, indirect inputs, electricity, depreciation cost, maintenance, and loss costs (in this paper, the real cost values are not furnished to preserve the company's business); $D_{r'p}$ is the annual demand of part $p$ for IF $r'$; $dc_{rr'}$ is the delivery cost from IF $r$ to IF $r'$; $no_{r'p}$ is the annual number of orders of part $p$ in IF $r'$; $ioc_{rp}$ is the internal order cost of part $p$ for IF $r$; $s$ is a supplier; $z_{sr'p}$ is a binary variable (1 if a route exists from supplier $s$ to IF $r'$ to deliver part $p$); $price_{sp}$ is the price of a unit of part $p$ from supplier $s$; and $sdc_{sr'}$ is the delivery cost from supplier $s$ to IF $r'$.

\subsubsection{Constraints}
The problem constraints and their interpretations are as follows: $i)$ the maximum number of 3D printers in a PC, $max^{P}$, must be lower than a limit inserted:
    \begin{align*}
      n_{r}\leq max^{P},\forall r \in R;
      \stepcounter{equation}\tag{\theequation}\label{eq:7}
    \end{align*}
$ii)$ the routes must be less than or equal to an input lead time, meaning that the production and delivery of a spare part must be performed in maximum hours of $max^{H}$ as
    \begin{align*}
      y_{rr'p}*\left(pt_{rp}+dt_{rr'}+ioct_{rp}\right) \leq max^{H},\forall r,r' \in R, p \in P,
      \stepcounter{equation}\tag{\theequation}\label{eq:8}
    \end{align*}
where $pt_{rp}$ is the processing time for a 3D printer to produce part $p$ in IF $r$, $dt_{rr'}$ is the delivery time (in hours) from IF $r$ to IF $r'$, and $ioct_{rp}$ is the internal order time (in hours) of part $p$ for IF $r$; $iii)$ an ES supply must be less than or equal to an input lead time (the maximum time from production to deliver a product), meaning that the acquisition and receiving of a spare part from an ES must be conducted in maximum hours of $max^{H}$:
    \begin{align*}
      z_{sr'p}*\left(st_{sr'}+oct_{sp}\right)\leq max^{H},\forall s \in S,r' \in R, p \in P,
      \stepcounter{equation}\tag{\theequation}\label{eq:9}
    \end{align*}
where $st_{sr'}$ is the delivery time (in hours) from supplier $s$ to IF $r$, and $oct_{sp}$ is the order time (in hours) of part $p$ of supplier $s$ to an IF; $iv)$ 3D printers should only be allocated to existent PCs as follows:
    \begin{align*}
      n_{r}\leq x_{r}*max^{P},\forall r \in R;
      \stepcounter{equation}\tag{\theequation}\label{eq:10}
    \end{align*}
$v)$ if a PC exists, it must supply as follows:
    \begin{align*}
      y_{rr'p}\leq x_{r},\forall r,r' \in R, p \in P;
      \stepcounter{equation}\tag{\theequation}\label{eq:11}
    \end{align*}
$vi)$ if an IF supplies a part, a route to another IF that receives it must exist as follows:
    \begin{align*}
      \sum_{r' \in R, p \in P} y_{rr'p}\geq x_{r},\forall r \in R;
      \stepcounter{equation}\tag{\theequation}\label{eq:12}
    \end{align*}
$vii)$ a route from a PC must only exist if there are demands:
     \begin{align*}
      y_{rr'p} \leq hasDemand_{r'p},\forall r,r' \in R, p \in P,
      \stepcounter{equation}\tag{\theequation}\label{eq:13}
    \end{align*}
where $hasDemand_{r'p}$ is a binary variable that indicates if an IF $r'$ demands part $p$; $viii)$ a route from an ES must only exist if there are demands:
    \begin{align*}
      z_{sr'p} \leq hasDemand_{r'p},\forall s \in S, r' \in R, p \in P;
      \stepcounter{equation}\tag{\theequation}\label{eq:14}
    \end{align*}
$ix)$ the demand must be supplied by a PC or an ES as follows:
    \begin{align*}
      \sum_{r \in R} y_{rr'p} + \sum_{s \in S}z_{sr'p} \geq hasDemand_{r'p},\forall r' \in R, p \in P;
      \stepcounter{equation}\tag{\theequation}\label{eq:15}
    \end{align*}
$x)$ the capacity must be sufficient to supply the demanded part (it should be noted that the capacity is half of the total required to grant print availability when an order arrives at a PC in a real time operation) as follows:
     \begin{align*}
      \sum_{r' \in R, p \in P} y_{rr'p}*D_{r'p}*pt_{rp} \leq n_{r}*\frac{C^{P}}{2},\forall r \in R,
      \stepcounter{equation}\tag{\theequation}\label{eq:16}
    \end{align*}
where $C^{P}$ is the annual capacity of a 3D printer (in hours); $xi)$ one and only one sender must supply the demanded entirely, whether it is an ES or a PC, as follows:
     \begin{align*}
      \sum_{r \in R} y_{rr'p}+ \sum_{s \in S} z_{sr'p} = 1,
      \\ \forall r' \in R, p \in P \mid r',p \in hasDemand_{r'p}.
      \stepcounter{equation}\tag{\theequation}\label{eq:17}
    \end{align*}

\subsubsection{Inputs}
The inputs of the optimization model include the pre-processed outputs from the $p$-median model (described in subsection \ref{subsec:preprocessing output}), annual fixed cost of a PC, maximum lead time, price of the 3D printer, capacity of the 3D printer, maximum number of printers in a PC, cost of producing a spare part, time spent producing a spare part, cost of supplying a spare part from an ES, suppliers' delivery time to supply IFs, suppliers' delivery costs by supplying IFs, and internal order cost.

\subsubsection{Setup}
This setup considers the dimensions (width, height, and depth) of the spare parts and the ZIP postal codes of the candidate IFs to obtain the costs and delivery time of the post office service. Thereafter, these data are employed to calculate the delivery costs to distribute the product. They are calculated using a web service through Brazil's post office, called the \textit{Correio} API\footnote{\url{https://pagseguro.uol.com.br/para-seu-negocio/online/envio-facil}}.

\subsubsection{Optimization model}
This is presented below, where its parameters, decision variables, constraints, and objective function are discussed.

\subsubsection{Parameters}
$r$, $p$, $s$, $D_{rp}$, $no_{rp}$, $FC^{R}$, $FC^{P}$, $C^{P}$, $max^{P}$, $max^{H}$, $pt_{rp}$, $cp_{rp}$, $uc_{rp}$, $price_{sp}$, $dc_{rr'}$, $dt_{rr'}$, $sdc_{sr}$, $st_{sr}$, $ioc_{rp}$, $ioct_{rp}$, $hasDemand_{rp}$, and $oct_{sp}$.

\subsubsection{Decision variables}
$x_{r}$, $y_{rr'p}$, $z_{srp}$, and $n_{r}$.

\subsubsection{Outputs}
These consist of the solutions of designing an SC using an AM process, selecting each IF that will become a PC or DC. This ensures that the demand will be satisfied by allocating some 3D printers to each AM facility. The outputs also determine the routes of a supplier of each spare part. Therefore, the outputs are PC data, routes, the capacity of the production (number of 3D printers), external suppliers, internal supply, and production.

It should be noted that the model assumes the rent to allocate a facility as an annual fixed cost; thus, costs to buy and install a PC are not considered. The selected location to the install a PC is addressed as a municipality range, thereby enabling some operational restrictions, such as room size, room temperature, noise, material handling, and disposal, to be neglected. A location that satisfies these restrictions is considered to exist in the selected municipality. In addition, for simplicity, the model does not consider some operational constraints such as the time spent on nesting parts in 3D printers, supplier of raw materials, spare part bounced back from certificate providers, post-treatment, and post-processors before the distribution phase.

\section{Experimental Results}
\label{sec:experimental_results}
In this section, a case study of an elevator maintenance service provider in Brazil, seeking to create an SC with AM units, is used to test and evaluate the proposed approach. In this sense, three problem scenarios are presented and tested. The experiments were performed on a PC with 8 GB of RAM and an Intel Core i5-3337U processor (1.8 GHz), and the algorithms were implemented using the software IBM ILOG CPLEX Optimization Studio.

\subsection{Data set description}
Each problem scenario considered five different types of spare parts.
The \textit{first scenario} had low demand, real data (87 demands/order lines), and 32 candidate cities; the \textit{second scenario} had medium demand, artificial data (1,000 demands/order lines), and 32 candidate cities; and the \textit{third scenario} addressed more geographic locations (98 candidate cities) and 267 artificial demands/orders, providing a scattered demand. The artificial data were created based on the application of the Monte Carlo method for the normal distribution of the first scenario.

\subsection{Parameter description}
\label{subsec:parameter_description}
In the first and second scenarios, the parameter was equated to $p = 32$, and it was $p$=$98$ for the third scenario. In all the scenarios, the other parameters were as follows:
$timeWeight$=$0.7$;
$distanceWeight$=$0.3$;
$FC^{P}$=$\$11,500.00$;
$FC^R$=$\$20,000.00$;
$C^{P}$=$2,112$ hours;
$max^{P}$=$5$ printers; and
spare part processing (printing) costs as part$_1$=\$22.00, part$_2$=\$7.00, part$_3$=\$21.00, part$_4$=\$14.00, part$_5$=\$47.00.
The model parametrization considered a Fullscale XT Plus printer with two extruders using a 1.75-mm$^2$ PLA filament and layer resolution of 7.62 mm to compose the measure of variables $uc_{rp}$ and $pt_{rp}$. The applied set of spare parts in this use case is presented in Fig. \ref{fig:set_of_spare_parts}.

\setcounter{figure}{2}
\begin{figure}[!htb]
\centering
\par\vspace{-0.3cm}\par
\includegraphics[width=0.5\textwidth]{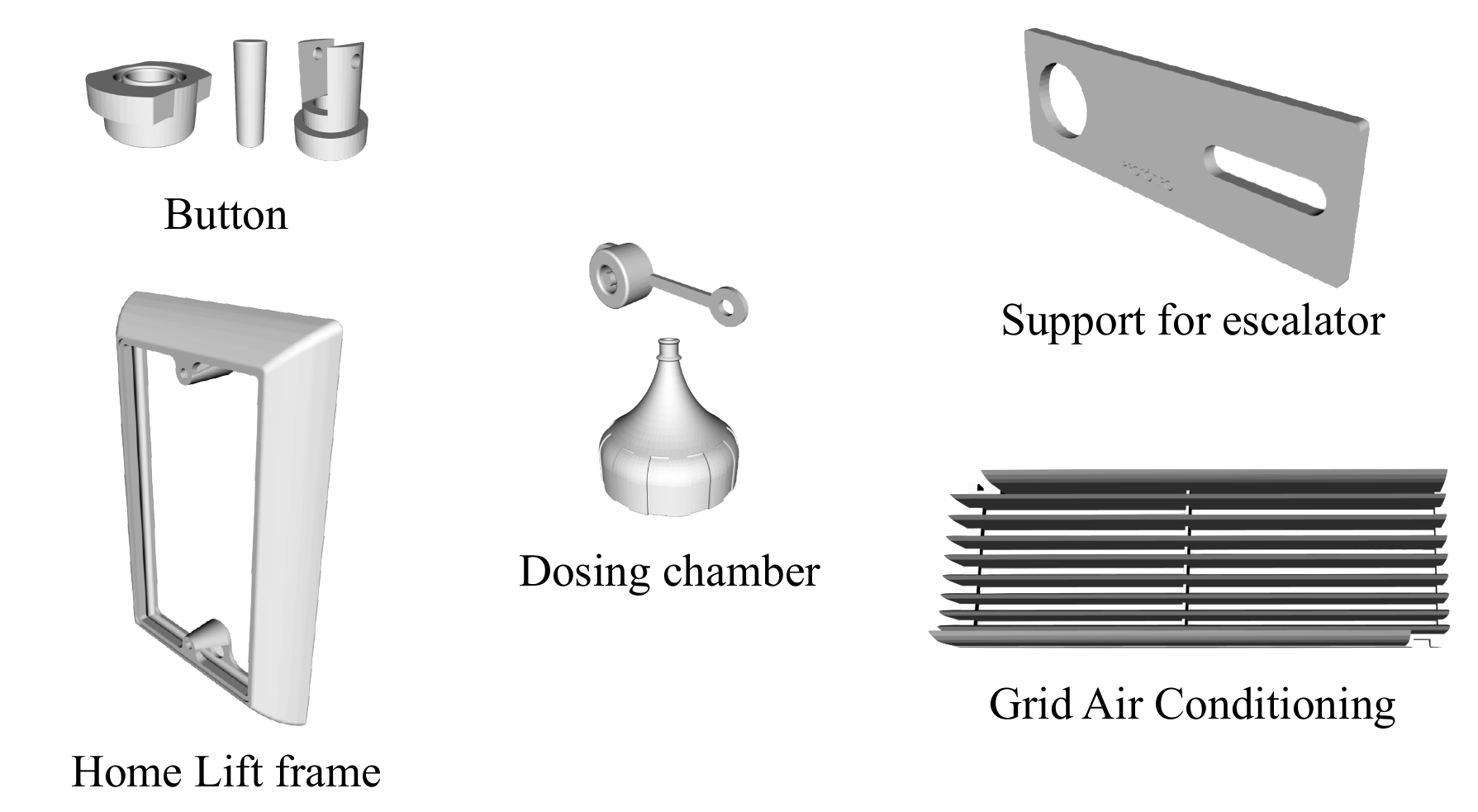}
\par\vspace{-0.3cm}\par
\caption{Set of 3D printed spare parts.}
\label{fig:set_of_spare_parts}
\end{figure}
\par\vspace{-0.6cm}\par

\subsection{Results and discussion}

\subsubsection{First scenario} low demand and real data

\begin{figure}[!htb]
\centering
\par\vspace{-0.5cm}\par
\includegraphics[width=0.5\textwidth]{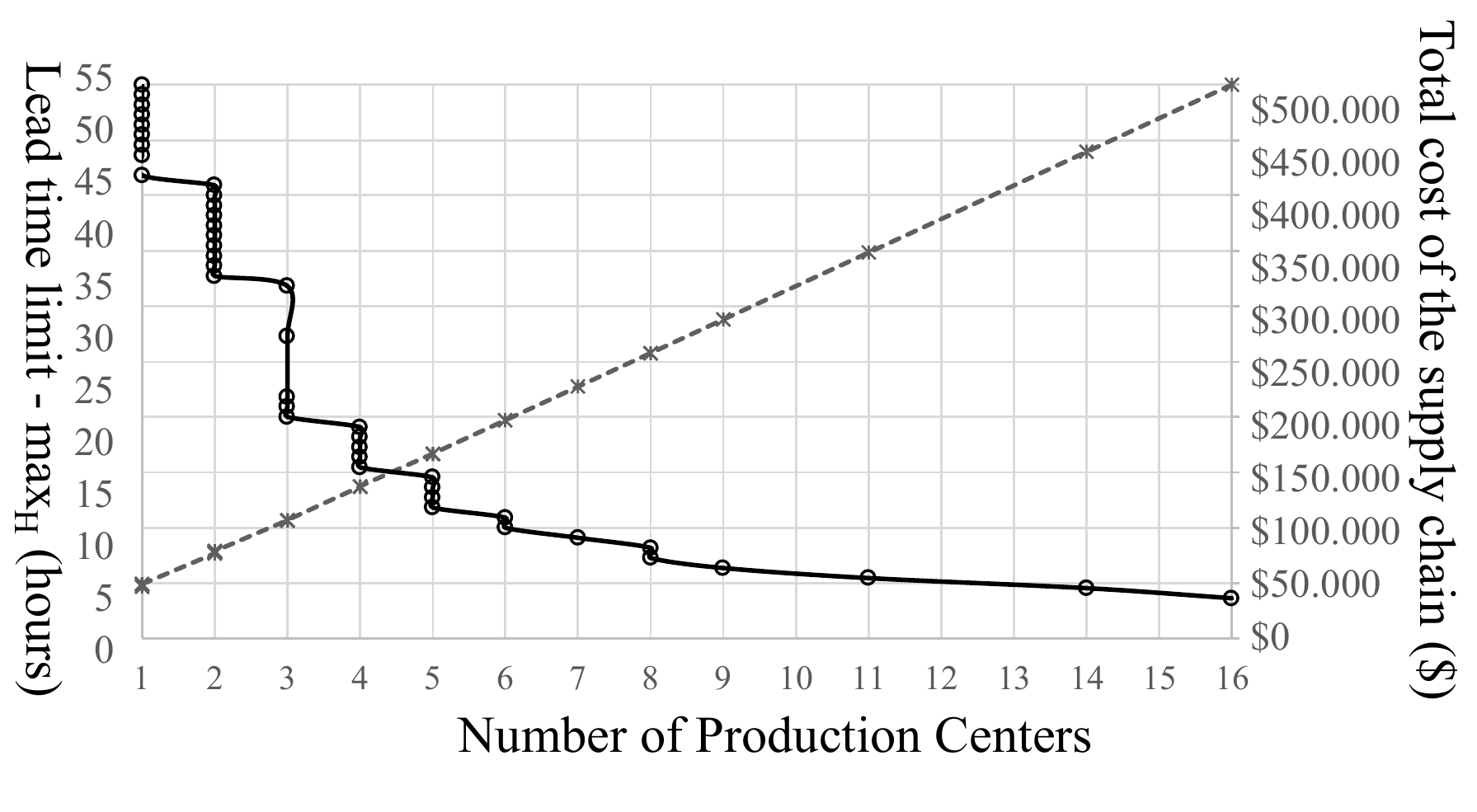}
\par\vspace{-0.5cm}\par
\caption{Scenario 1 - Results of the cost {\color{darkgray} \textbf{\huge \--\--}} and the lead time {\color{black} \textbf{\huge \textendash}} when the number of PCs increases.}
\par\vspace{-0.3cm}\par
\label{fig:results}
\end{figure}

To evaluate the effect of constraining the lead time in the generated solutions, tests were executed by the varying parameter $max^{H}$ from 4 to 55 h. Experiments have demonstrated that a solution becomes infeasible under 4 h, whereas, over 55 h, the results become invariable. Fig. \ref{fig:results} shows the obtained results of the first scenario, where \textit{Set lead time} is associated with the $max_H$ parameter, \textit{Total cost of the SC} is the total cost of redesigning the SC to satisfy the demands, and the \textit{Number of Production Centers} is the number of PCs (given by the sum of the values of decision variables of type $x_{r}$). 

The obtained results indicated that when the lead time limit was $55$ h, the obtained number of PCs was $1$, and the total cost of the SC was $\$48,402.89$ with one 3D printer and 2 d, 6 h, and 22 min to supply the most distant demand. Moreover, when the lead time limit was $4$ h, the obtained number of PCs was $16$ and the total cost of the SC was $\$519,289.98$ with each PC with one 3D printer and 4 h to supply the most distant demand. Moreover, the results indicated that a decrease in the $max^{H}$ value implied an increase in the number of PCs and total cost. The best cost-benefit SC design may be expressed graphically in the intersection of the two lines in Fig. \ref{fig:results}, which results in 4 PCs, with each PC having one 3D printer, and the total cost of the SC is $\$142,708.18$ with 17 h to supply the most distant demand.

\subsubsection{Second scenario} medium demand and artificial data

\begin{figure}[!htb]
\centering
\par\vspace{-0.5cm}\par
\includegraphics[width=0.5\textwidth]{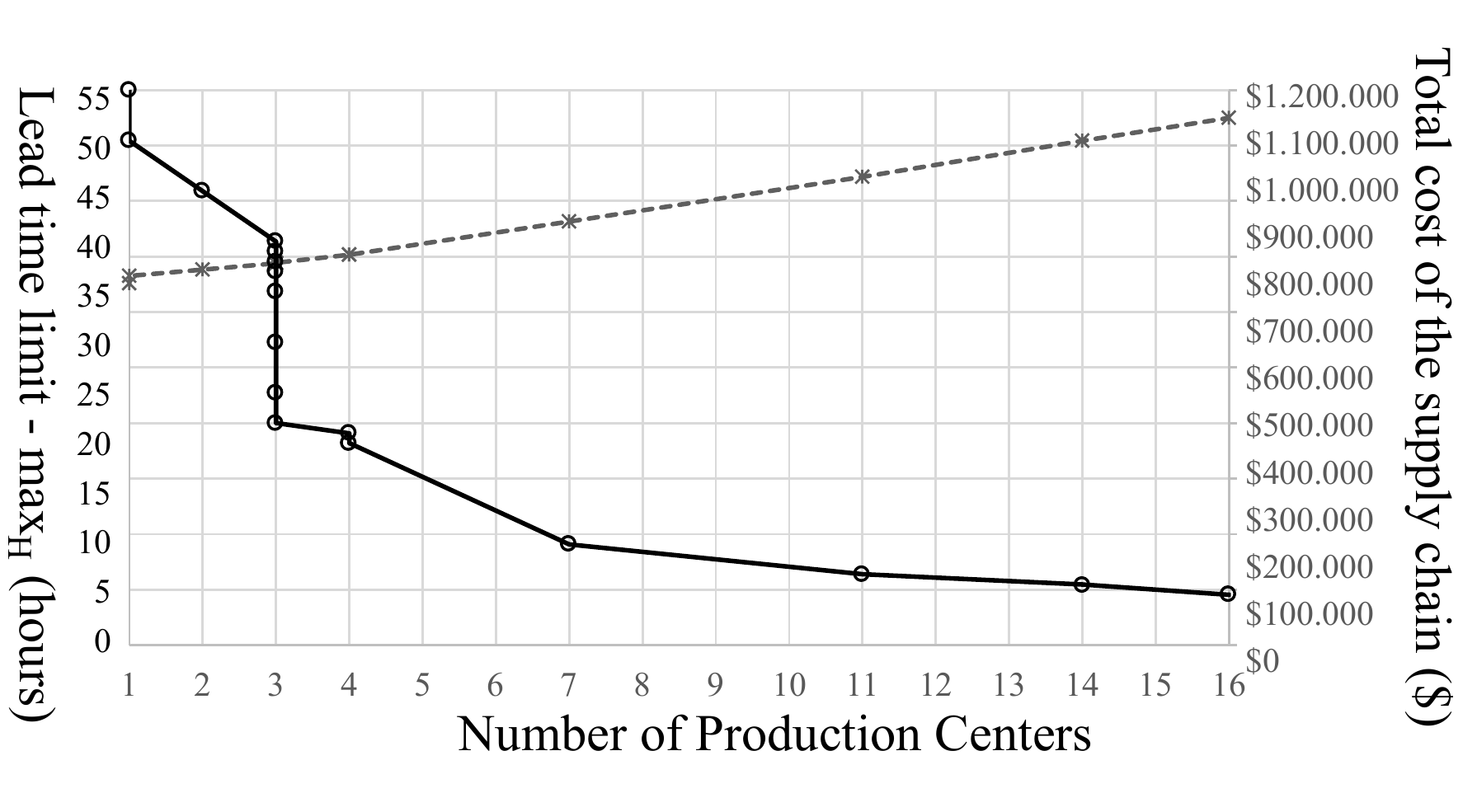}
\par\vspace{-0.5cm}\par
\caption{Scenario 2 - Results of the cost {\color{darkgray} \textbf{\huge \--\--}} and the lead time {\color{black} \textbf{\huge \textendash}} when the number of PCs increases.}
\label{fig:results1000}
\par\vspace{-0.3cm}\par
\end{figure}

The tests were performed by varying $max_H$ from 4 to 55 h, as explained above. Fig. \ref{fig:results1000} depicts the obtained results of the second scenario. The obtained results indicated that when the lead time limit was $55$ h, the obtained number of PCs was $1$, and the total cost of the SC was $\$816,221.83$ with one 3D printer and 2 d, 6 h, and 22 min to satisfy the most distant demand. Moreover, when the lead time limit was $4$ h, and the number of PCs was $18$, the total cost of the SC was $\$1,252,495.00$, each PC with one 3D printer and 4 h to satisfy distant demands. In the second scenario, the results also indicated that a decrease in the $max^{H}$ value implied an increase in the number of PCs and total cost. The best solution may be expressed graphically in the intersection of the two lines in Fig. \ref{fig:results1000}, which is produced for 4 PCs, and the total cost of the SC is $\$816,682.03$ with 1 d and 13 h to satisfy the most distant demand.

\subsubsection{Third scenario - low scattered demand and artificial data}

To evaluate the effect in a scenario with scattered demand (more candidate cities), tests were performed by considering the 98 most populous cities in Brazil \cite{ibge2017estimativas} as candidate cities. This decision was based on the assumption that the most populous cities have more elevators and they require spare parts. The selected cities have more than 270,000,000 inhabitants.

\begin{figure}[!htb]
\centering
\par\vspace{-0.4cm}\par
\includegraphics[width=0.5\textwidth]{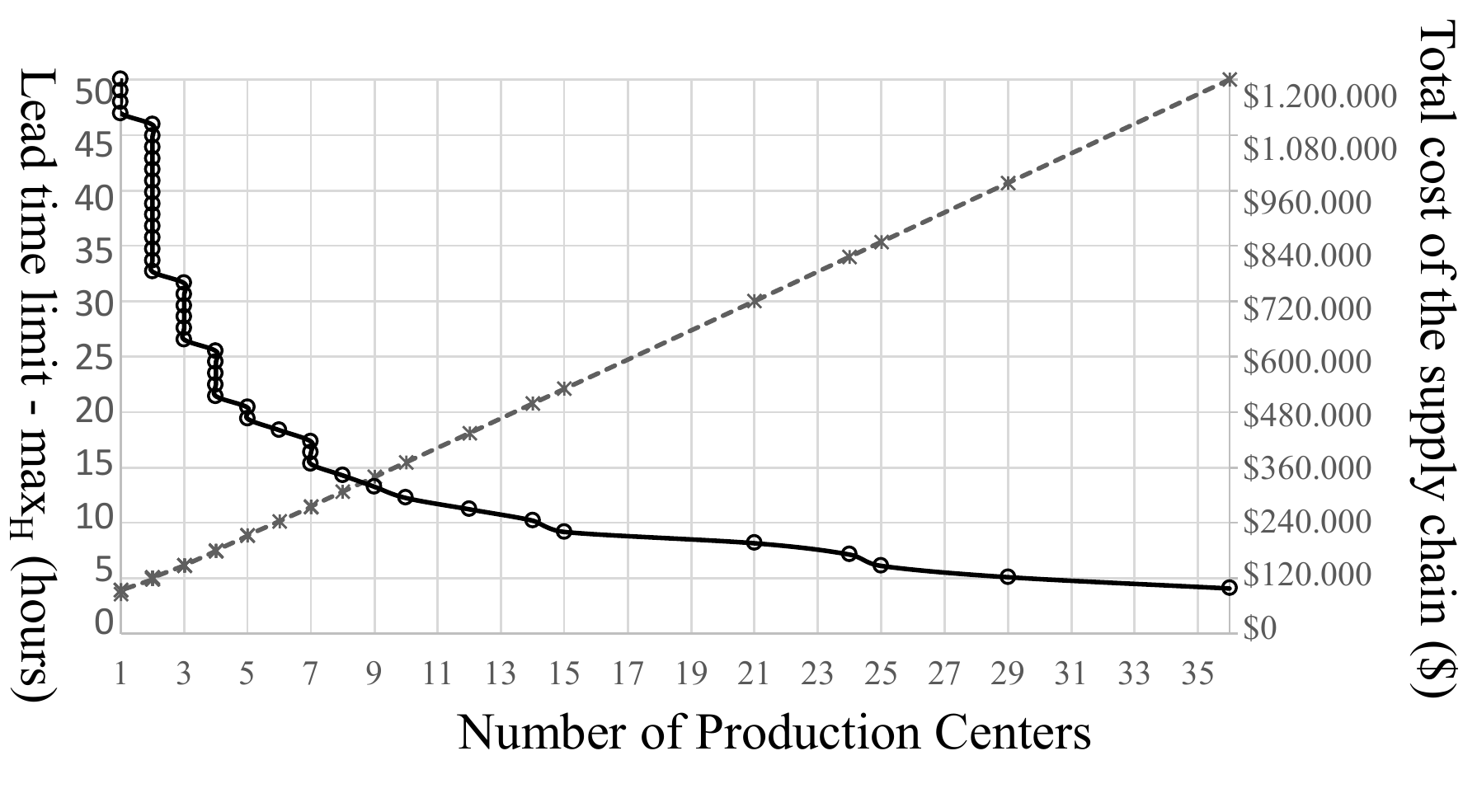}
\par\vspace{-0.6cm}\par
\caption{Scenario 3 - Results of the cost {\color{darkgray} \textbf{\huge \--\--}} and the lead time {\color{black} \textbf{\huge \textendash}} when the number of PCs increases.}
\label{fig:results_cenario_3}
\par\vspace{-0.1cm}\par
\end{figure}

As conducted in the previous scenarios, tests were performed by varying the parameter $max^{H}$ to evaluate the effect of restricting the lead time of the generated solutions. Experiments have demonstrated that a solution becomes infeasible under 4 h and results become invariable over 49 h. Fig. \ref{fig:results_cenario_3} shows the obtained results of the third scenario.

The obtained results revealed that when the lead time limit was $49$ h, the obtained number of PCs was $1$, and the total cost of the SC was $\$84,146.83$ with one 3D printer and 2 d, 23 h, 40 min, and 30 s to supply the most distant demand. Moreover, when the lead time limit was $4$ h, the obtained number of PCs was $36$, and the total cost of the SC was $\$1,180,964.73$ with each PC having one 3D printer and 4 h to supply the most distant demand.

\begin{figure}[!htb]
\centering
\includegraphics[width=0.5\textwidth]{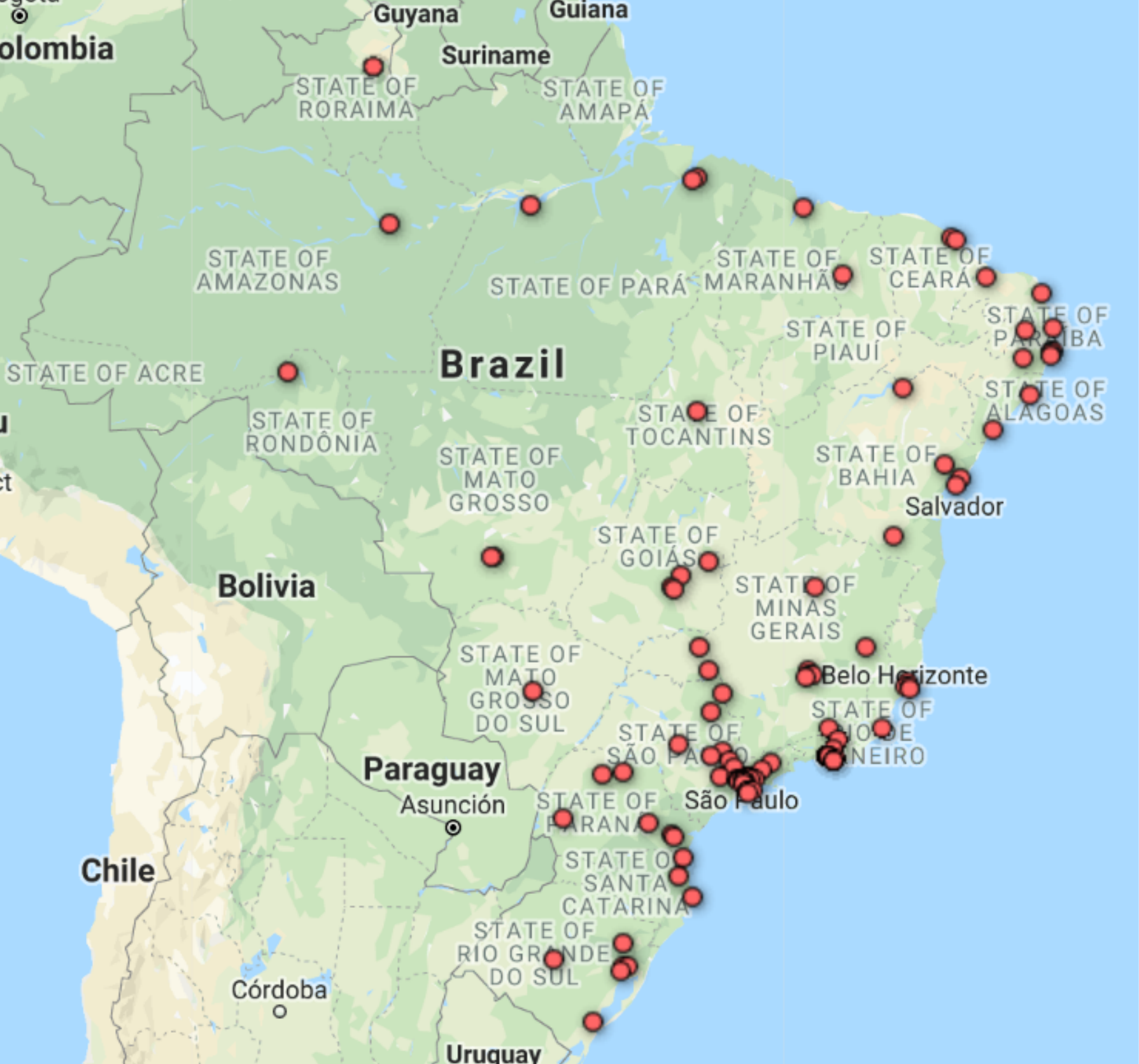}
\par\vspace{-0.3cm}\par
\caption{Scenario 3 - Map with the 98 candidate cities.}
\label{fig:map_input_cenario_3}
\par\vspace{-0.3cm}\par
\end{figure}

\begin{figure}[!htb]
\centering
\includegraphics[width=0.5\textwidth]{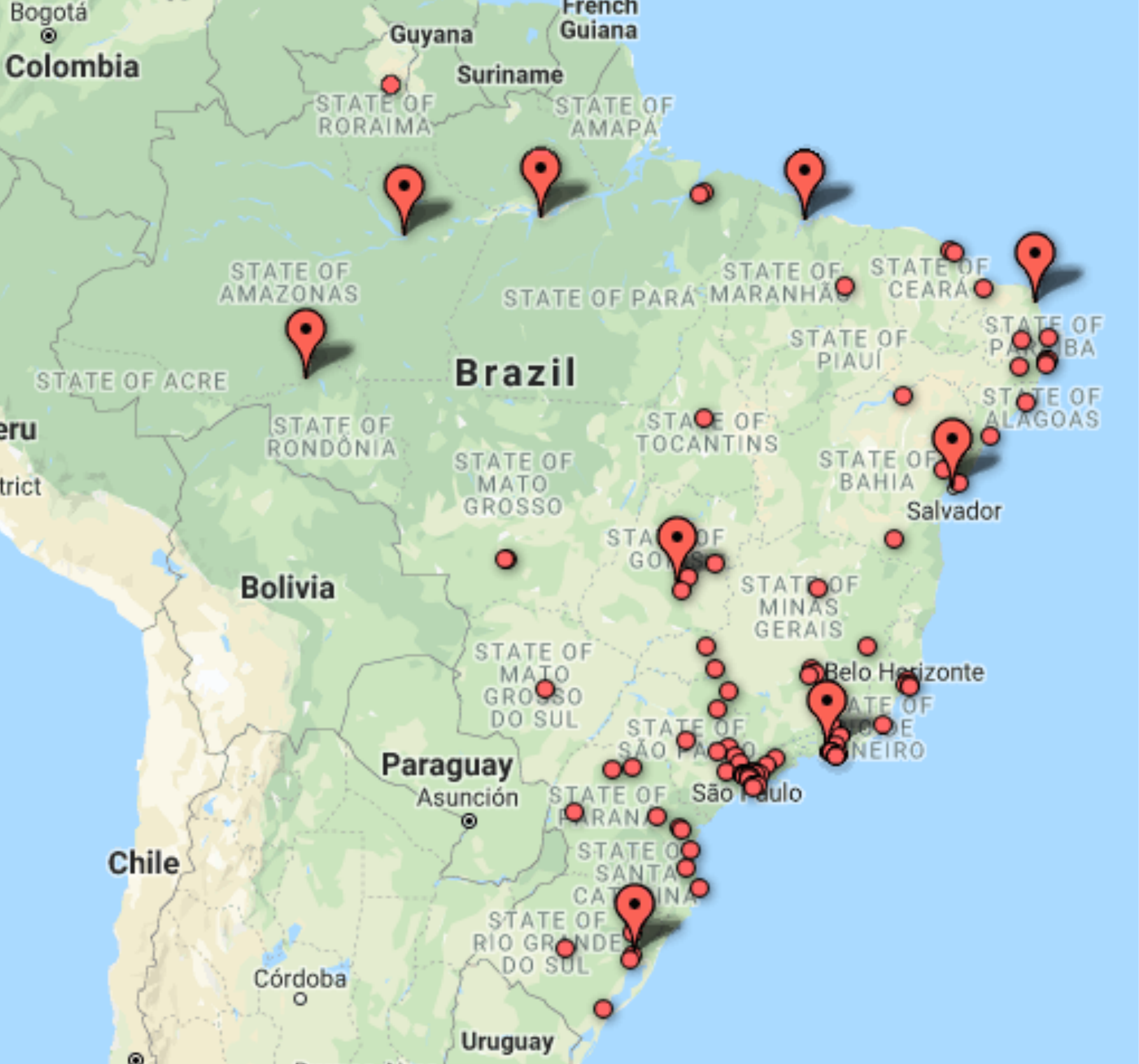}
\par\vspace{-0.3cm}\par
\caption{Scenario 3 - The selected cities/PCs, represented by large tags, in the best cost-benefit result.}
\label{fig:map_output_cenario_3}
\end{figure}

The best cost-benefit SC design may be expressed graphically in the intersection of the two lines in Fig. \ref{fig:results_cenario_3}, which results in 9 PCs with each PC having one 3D printer; the total cost of the SC is $\$333,624.43$ with 13 hours to supply the most distant demand. To show the best cost-benefit solution, we created Figs. \ref{fig:map_input_cenario_3} and \ref{fig:map_output_cenario_3}, which depict the 98 candidate cities and selected cities/PCs (represented by large tags), respectively. The results revealed that the optimization algorithm distributed PCs among the scattered candidate cities to minimize the distribution costs.

Table \ref{table:Results} indicates that optimal solutions were determined by the proposed model in a short time for scenarios 1 and 2 and medium time for scenario 3. The average execution times to determine an optimal solution was $48.3$, $127.93$, and $519.7$ s for scenarios 1, 2, and 3, respectively. In large demand scenarios, the model could not determine solutions in a short time, making adopt other strategies, such as heuristics or meta-heuristics, important to accomplish better results in less time for such scenarios.

\setlength{\textfloatsep}{1pt}
\begin{table}[!ht]
\par\vspace{-0.5cm}\par
\captionsetup{belowskip=0pt,aboveskip=0pt}
\caption{Performance results of the proposed approach}
\par\vspace{-0.3cm}\par
\label{table:Results}
{\scriptsize
\begin{center}
\begin{tabular}{cccc}
\toprule
Problem scenario & Number of IFs & Average execution time (seconds) \\\midrule
Scenario 1 & 32 & 
48.3 \\\midrule
Scenario 2 & 32 & 127.93 \\\midrule
Scenario 3 & 98 & 519.7 \\\midrule
\end{tabular}\\
\par\vspace{-0.5cm}\par
\end{center}}
\end{table}

Note that a lead time limit (composed by the production time and the travel time) exists even if the number of PCs is large. If the number of PCs is large and each IF becomes a PC, the travel time is zero, and the lead time is the production time of the spare part. Because the maximum production time is 4 h, the lead time limit is also 4 h.

\section{Conclusions}
\label{sec:conclusions}
This paper proposes a design approach for AM SCs. The approach was tested by considering a spare part SC using an MTO strategy. Here, the main objective was to optimize the total costs of producing and delivering spare parts (produced by 3D printers) for a given lead time. The proposed optimization model, which uses a $p$-median model and a location-allocation problem, was solved using MILP. A real-world use case considered a provider of elevator maintenance services. The experiments revealed that the proposed model can determine optimal solutions within an acceptable time of execution. Moreover, the results revealed the promising capabilities of the proposed model to address the new SC design challenges from the forthcoming widespread use of 3D printers in manufacturing SCs.

Future studies can be devoted to the application of the proposed design approach to other real-world scenarios. In addition, further research can involve the development of optimization models that can produce optimal solutions in more complex design instances of AM SCs. Additionally, the use of heuristic approaches can be investigated to solve high demand scenarios. Finally, another course of relevant research is the full exploitation of cyber-physical systems to support the operational allocation of production orders to AM facilities in the designed manufacturing SCs.

\section*{Acknowledgment}
This project has received funding from the European Union’s Horizon 2020 research and innovation programme and from the Brazilian Ministry of Science, Technology and Innovation through Rede Nacional de Pesquisa under the Grant Agreement 777096.

\bibliographystyle{IEEEtran}
\bibliography{TII-19-4426.bib}
\vspace{-3cm}
\begin{IEEEbiography}[{\includegraphics[width=1in,height=1.25in,clip,keepaspectratio]{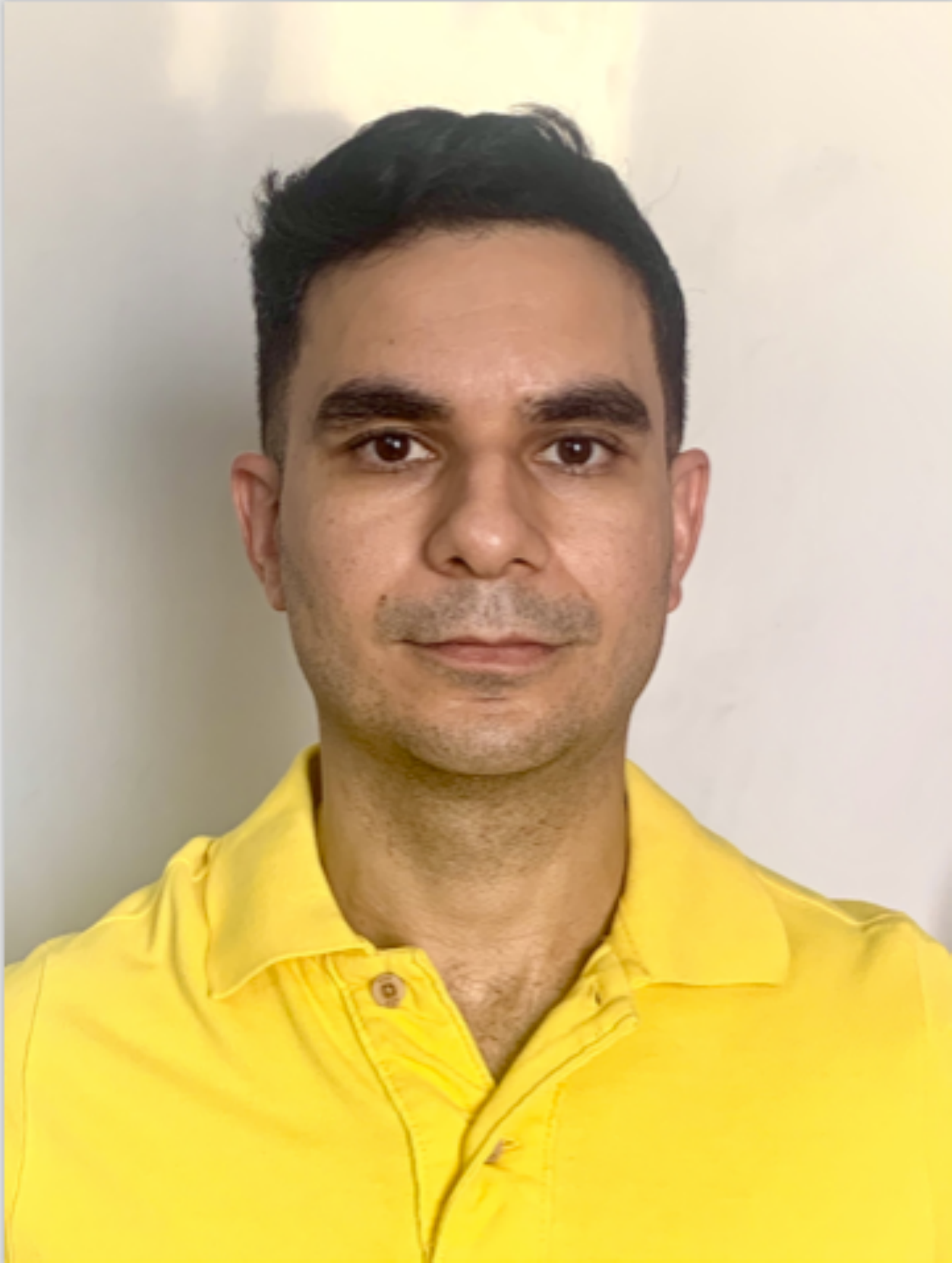}}]{Filipe Marinho de Brito} is currently pursuing a Ph.D. degree in Computer Engineering at the School of Electrical, Mechanical, and Computer Engineering, Federal University of Goiás (EMC/UFG-GO), Brazil. M.Sc. in Computer Engineering (2013), B.Sc. in Public Security Management Fire Brigade Square by the State University of Goiás (2006), and B.Sc. in Computer Engineering at the Federal University of Goiás (2008). Currently Systems Analyst, Business Intelligence Analyst, and Developer at Fire Department of Goiás and a Researcher on the Flexible and Autonomous Manufacturing Systems for Custom-Designed Products project (FASTEN). Research interests include Optimization Techniques, Microservices Systems, Distributed Computing, Information Technology, and Industrial Systems.
\end{IEEEbiography}
\vspace{-3cm}
\begin{IEEEbiography}[{\includegraphics[width=1in,height=1.25in,clip,keepaspectratio]{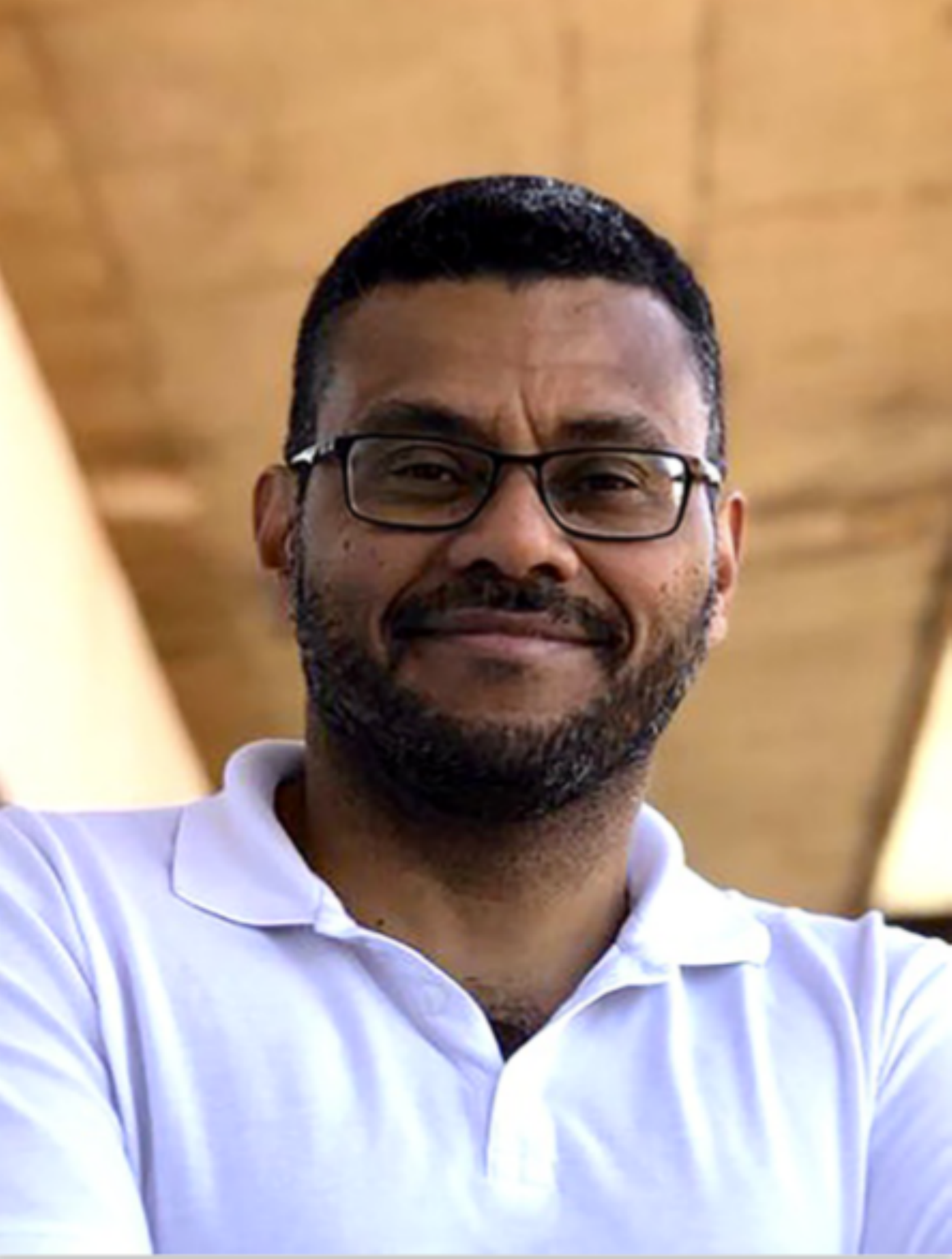}}]{Gélson da Cruz Júnior} holds B.Sc. degree in Electrical Engineering - Ilha Solteira from Paulista State University Júlio de Mesquita Filho (1990), M.Sc. in Electrical Engineering from the State University of Campinas (1994) and a Ph.D. in Electrical Engineering from the State University of Campinas (1998). He held a post-doctorate at INESC-Porto between 2006 and 2007. He is currently a full professor and vice-coordinator of the postgraduate course in Electrical and Computer Engineering at the School of Electrical, Mechanical and Computing Engineering of the Federal University of Goiás (EMC/UFG-GO). He has experience in the area of Electrical Engineering, with an emphasis in Systems Engineering, working mainly on the following topics: neural networks, operational research, forecasting models, fuzzy logic and planning.
\end{IEEEbiography}
\vspace{-3cm}
\begin{IEEEbiography}[{\includegraphics[width=1in,height=1.25in,clip,keepaspectratio]{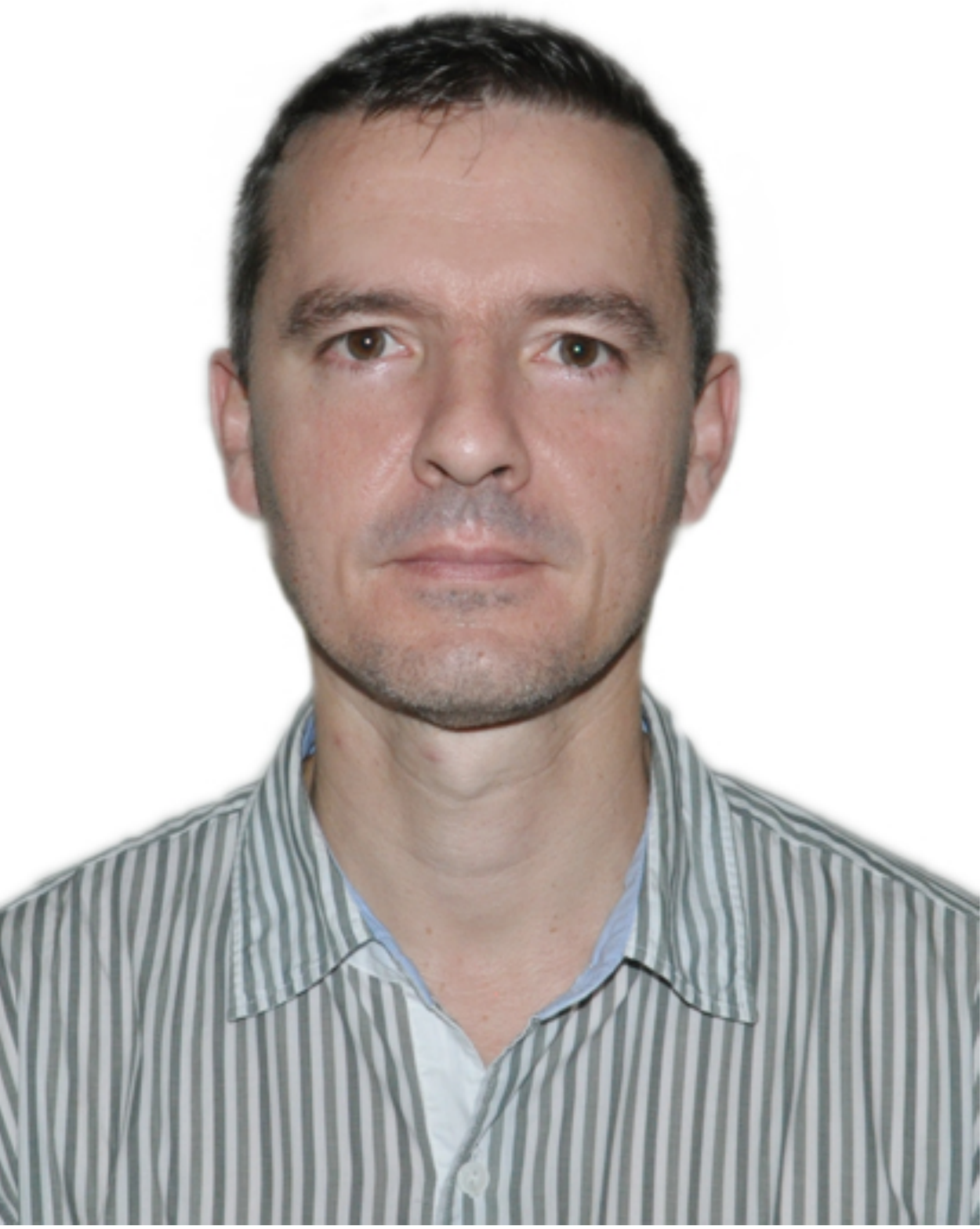}}]{Enzo Morosini Frazzon} (Dr.-Ing. from the University of Bremen, 2006 - 2009) is Associate Professor at the Department of Production Engineering and Systems (EPS) of the Federal University of Santa Catarina (2010 - ...). He was a Postdoctoral Researcher at the BIBA/Bremen (2009 - 2010) and Visiting Scholar at the University of Parma (2018 - 2019). Research Fellow of INESC Brasil. He leads projects on the: integration of production and logistic processes; intelligent production and logistic systems; Supply Chains; Optimization; Simulation; Data Analytics; Cyber-physical Systems; Advanced Manufacturing and Industry 4.0. He authored 100+ papers in international scientific conferences and journals.
\end{IEEEbiography}
\vspace{-3cm}
\begin{IEEEbiography}[{\includegraphics[width=1in,height=1.25in,clip,keepaspectratio]{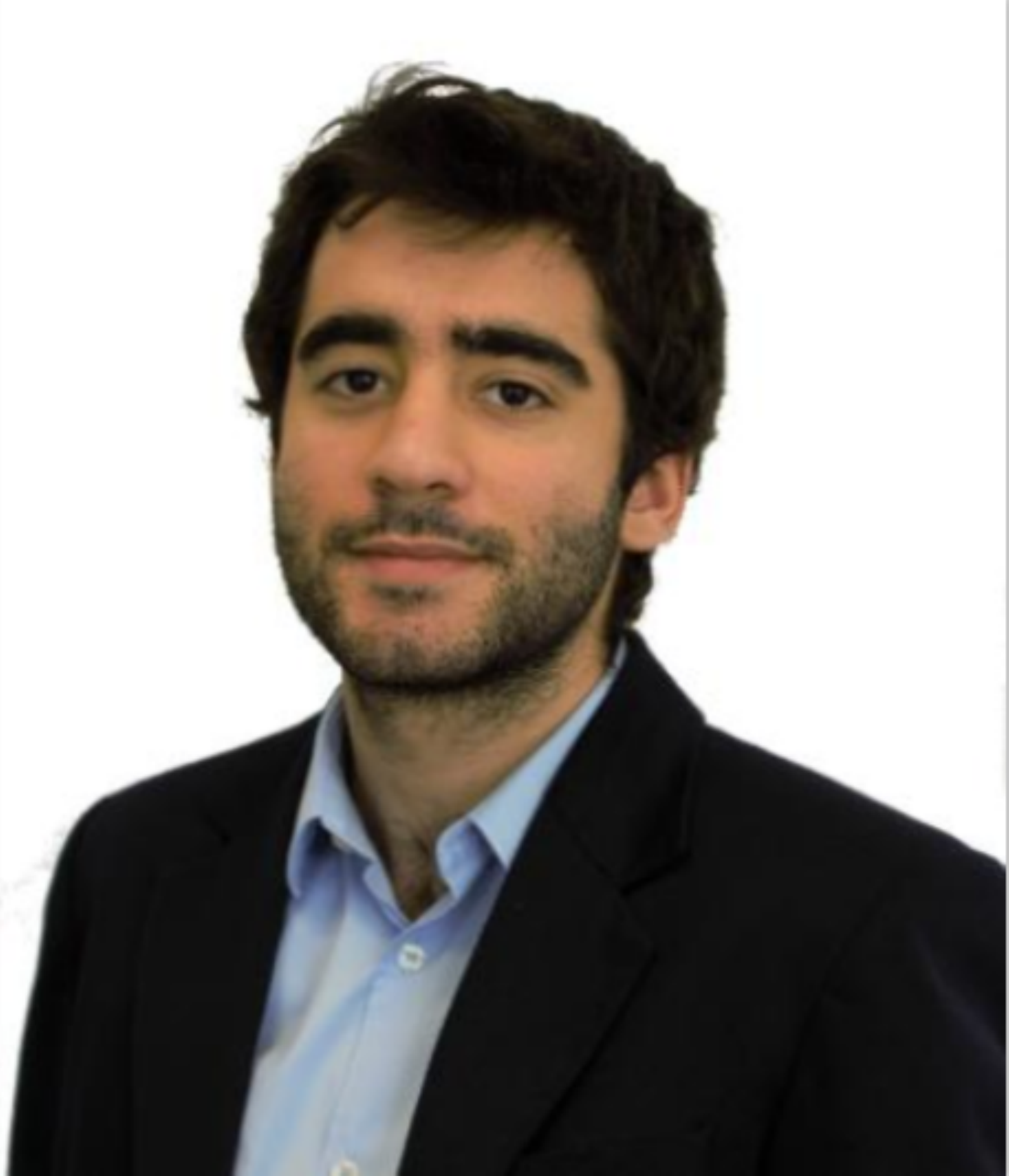}}]{João Pedro Tavares Vieira Basto} is Invited Assistant Professor at the Faculty of Engineering of the University of Porto (FEUP) and a researcher at INESC TEC, in Porto. He received his Masters degree in Electrical and Computers Engineering from FEUP in 2016, majoring in Automation and specializing in Industrial Management. After graduating, he worked for an analytics-consulting firm, where he was involved in a wide range of industrial management projects, in areas from stock management and demand forecasting to production scheduling. Afterwards, he joined INESC-TEC, where he has been leading research efforts on using simulation-optimization for supply chain management and production balancing and scheduling.
\end{IEEEbiography}
\vspace{-3cm}
\begin{IEEEbiography}[{\includegraphics[width=1in,height=1.25in,clip,keepaspectratio]{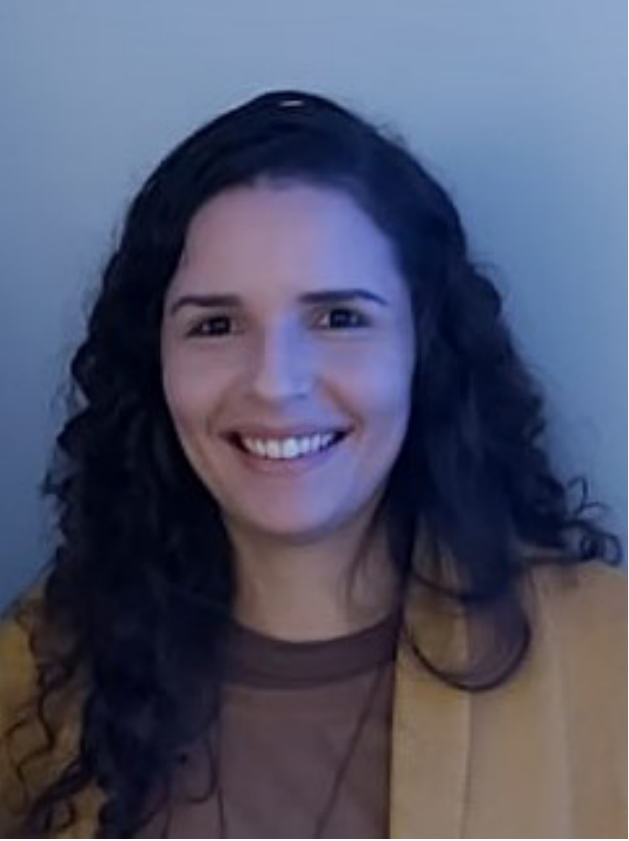}}]{Symone Gomes Soares Alcalá} is currently an Adjunct Professor at the Federal University of Goiás, Faculty of Sciences and Technology, Production Engineering. She received her Ph.D. degree in Electrical and Computer Engineering from the University of Coimbra, Portugal, in 2015; and her B.Sc. degree in Computer Engineering from the Pontifical Catholic University of Goiás (PUC-GO), Brazil, in 2009. She was a researcher at the Portuguese Institute of Systems and Robotics (ISR-Coimbra, University of Coimbra, Portugal) from 2010 until 2015, working in the field of Computational Intelligence Modeling for Industrial Processes. Her research interests are in Machine Learning, Optimization Techniques, Computational Intelligence Modeling, and Industrial Systems.
\end{IEEEbiography}

\end{document}